\documentclass[a4paper,20pt]{article}
    \usepackage[T1]{fontenc}
    \usepackage{graphicx}
    \usepackage{epsfig}
    \usepackage{amsmath}
    \usepackage{amsfonts}
    \usepackage{url}
    \renewcommand{\abstract}{}
\begin{document}
\makeatletter
\renewcommand{\@oddhead}{\textit{YSC'17 Proceedings of Contributed Papers} \hfil
\textit{M.G. Malygin, D.A. Iakubovskyi}}
\renewcommand{\@evenfoot}{\hfil \thepage \hfil}
\renewcommand{\@oddfoot}{\hfil \thepage \hfil}
\fontsize{11}{11} \selectfont

\title{Search for cyclotron absorptions from magnetars in the quiescence with
\textit{XMM-Newton}}
\author{\textsl{M.G. Malygin$^{1}$, D.A. Iakubovskyi$^{1,2}$}}
\date{}
\maketitle
\begin{center} {\small $^{1}$Kyiv National Taras Shevchenko University,
Glushkova ave., 2, 03127, Kyiv, Ukraine\\
$^{2}$Bogolyubov Institute for Theoretical Physics, 14-b, Metrolohichna str.,
Kiev, 03680, Ukraine\\
kolkosm@ukr.net}
\end{center}

\begin{abstract}
  In this work, we perform the detailed analysis of absorption features in
spectra of magnetar candidates observed by XMM-Newton satellite. No significant
line-like feature has been found. This negative result may indicate the
possible presence of smoothing out the absorption features mechanisms.
\end{abstract}

\section*{Introduction}

  Magnetars are strongly magnetized neutron stars powered with a super-strong
magnetic field (see e.g. \cite{mereghetti08,TW06} for a review). There are two
classes of objects which properties could be described in terms of the magnetar
model: Anomalous X-ray Pulsars (AXPs) and Soft Gamma Repeaters (SGRs). 
It is generally assumed that such sources spin down by magnetic dipole
radiation. The inferred  values of the surface magnetic field
strength $10^{14}-10^{15}$~G are larger than the so-called \textit{critical}
magnetic field $B_{crit} \equiv m_e^2c^3/e\hbar = 4{.}4\cdot10^{13}$~G 
where the effects of quantum electrodynamics should be taken into account. This
makes magnetars to be the unique natural laboratories to test such superstrong
magnetic fields.

  However, no direct measurements of magnetic fields in magnetars have been
presented so far. It was suggested in \cite{TLK02} to search for proton
cyclotron harmonic, which should appear in the x-ray band (2-10~keV) due to the
presence of the strong magnetic field:

\begin{equation}
\label{pc}
\hbar\omega^{cycl}_{p} = 0.63\left( 1 + z \right)^{-1} \left(
\frac{B}{10^{14} G} \right) \mbox{ keV.}
\end{equation}

  Here $\hbar\omega^{cycl}_{p}$ denotes the \textit{observed} energy of the
cyclotron feature, $\left(1 + z\right)^{-1} = \left(1-2GM/rc^2\right)^{1/2}$ is
the gravitational redshift ($\sim 0{.}8$ for a typical neutron star), $B$ is
the magnetic field strength. Previous search revealed some features during
flares and high states, which could be treated as cyclotron features, in
the spectra of 7 among 19 magnetar candidates (see Table~\ref{tab1}).

  Summarizing the results of Table~\ref{tab1}, we note that the most of the
absorption features were mainly observed during flaring stages and with
moderate-resolution instruments such as RXTE/PCA. It was also announced about
the absence of the absorption features in the \textit{quiescent} spectra of AXP
4U~0142+614,~\cite{Rea07b}, although they were predicted in earlier theoretical
works~\cite{HoLai01,Zane01}. The aim of this paper is to provide more extended
search for cyclotron features in the quiescent spectra of magnetars. For such
an analysis, a combination of good energy resolution and large effective area
is necessary. Therefore, we concentrate on EPIC cameras onboard
\textit{XMM-Newton} satellite, extending the earlier results
of~\cite{Rea07b,Oosterbroek04}, where the same instruments were used.

\section*{The Method and Results}

In this work, we used all publically available XMM-Newton/EPIC observations with
all three cameras (e.g. MOS1, MOS2 \& PN) in the \textit{Imaging} mode. It was
done to ensure the better statistics and to prevent possible calibration
uncertainties which may occur in \textit{Timing} mode. All event lists were
cleaned from soft proton solar flares using the standard
routine~\url{http://www.sr.bham.ac.uk/xmm2/xmmlight_clean.csh} v.~3.3. Source
and background regions were substracted manually from the image. Pile-up
checking was carried out with the \texttt{epatplot} procedure. If the predicted
ratios of the single-to-double events were not equal to 1 within the
\texttt{epatplot} error interval, such observations were not included in the
subsequent analysis. After taking all reductions 41 high-quality
observations for 12 objects remained.

\begin{table}
\begin{center}

\caption{Parameters of the previously detected absorption-like features found in
magnetar candidates. 'MC' denotes Monte Carlo simulations; \texttt{gauss} or
\texttt{cyclabs} denotes the \texttt{Xspec} model describing the absorption
feature.}

\begin{tabular}{p{28mm}p{13mm}p{28mm}p{23mm}p{39mm}p{14mm}}

\hline
\hline
Object & Energy, keV & Significance, method& Instrument &
Notes & References \\
\hline
1E~2259+586 & 5, 10 & -- & GINGA/LAC & during flux increase & \cite{IwKo92}\\
SGR~1806-20 & 5.0, 7.5, 11.2, 17.5 & 3.3$\sigma$, \texttt{gauss}, F-test (for a
set of features)& RXTE/PCA & in the harder part of a precursor &
\cite{Ibrahim02},
\cite{Ibrahim03}$^1$\\
4U~0142+614 & 4, 8, 14 & -- & RXTE/PCA & emissions, in the most energetic among
a sequence of bursts & \cite{Gavriil07}\\
1E~1048-5937 & 14 & 3.9$\sigma$, \texttt{gauss}, MC & RXTE/PCA & emission, in a
burst & \cite{Gavriil02}\\
            & 13 & 3.3$\sigma$, \texttt{gauss}, MC & RXTE/PCA & emission, at one
part of a bursts tail  & \cite{Gavriil06}\\
XTE~J1810-197$^2$ & 12.6 & 4.5$\sigma$, \texttt{gauss}, MC & RXTE/PCA &
emission in a burst tail & \cite{Woods05}\\
1RXS~J1708-4009 & 8.1 & 2.95$\sigma$, \texttt{cyclabs}, MC & \textsl{Beppo}SAX
 & the longest observation (200ks), during rising phase &
\cite{Rea07a}, \cite{Rea03}, \cite{Oosterbroek04}\\
SGR~1900+14 & 6.4$^3$ & 3.7$\sigma$, \texttt{gauss}, F-test & RXTE/PCA
& during precursor to the main burst & \cite{Strohmayer00} \\
\hline
\hline
\cline{1-1}
\multicolumn{6}{p{165mm}}{$^1$\rule{0pt}{11pt} In this work there is a significance estimation of a 5 keV
feature (3$\sigma$, C-statistic), which appeared in a variety of bursts, whereas
the set of features appeared only in one bursting episode of the long
precursor.}\\
\multicolumn{6}{p{165mm}}{$^2$\rule{0pt}{11pt} It was also reported about an absorption-like feature around
1.1~keV in the XMM-Newton/EPIC spectra, which has more readily been interpreted
as an absorption edge than a cyclotron absorption, however (see~\cite{Bernardini09}).}\\

\multicolumn{6}{p{165mm}}{$^3$\rule{0pt}{11pt} There was also a weak excess near 13~keV, but the authors found
it to be insignificant and interpreted the line as Fe K-$\alpha$ emission.}\\

\end{tabular}
\label{tab1}
\end{center}
\end{table}

\begin{figure}[!h]
\centering
\epsfig{file = 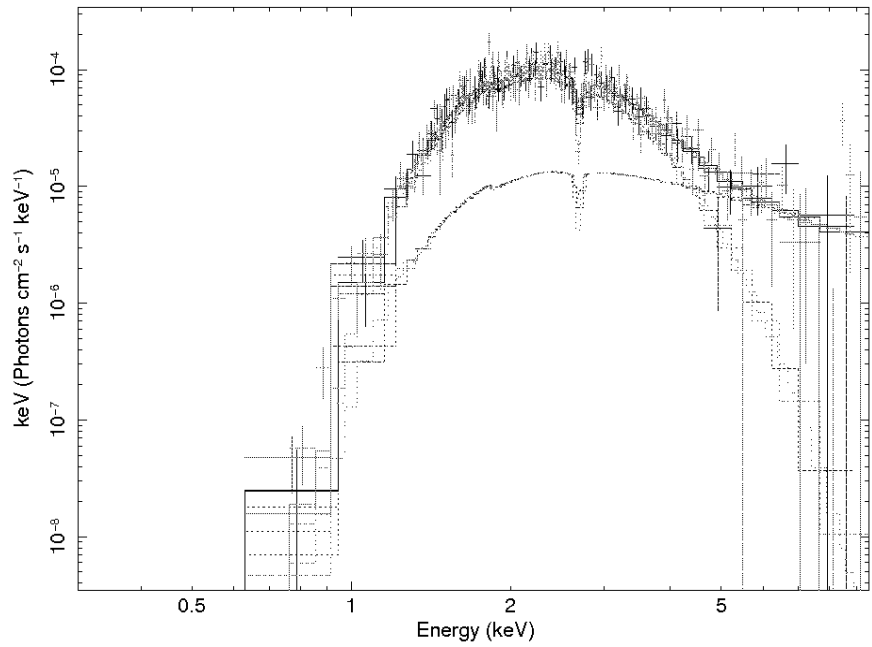, width = .72\linewidth}
\caption{The spectrum of 1E~1547.0-5408 with the highest significance 
of a cyclotron absorption of 1.94$\sigma$ estimated from the Monte Carlo
simulations. The data are fitted with model \texttt{phabs*(blackbody+powerlow)*cyclabs}
(see page~\pageref{MandR}).}
\label{fig1}
\end{figure}

\begin{figure}[!h]
\centering
\epsfig{file = 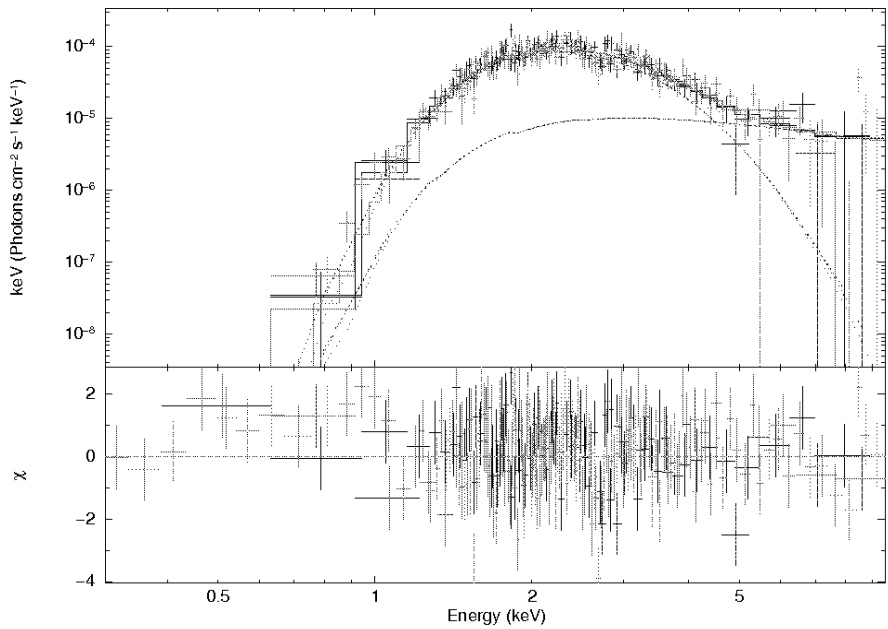, width = .73\linewidth}
\caption{The same as in Fig.~1 but without adding a cyclotron absorption.
The model residuals are shown in bottom part of the figure.}
\label{fig2}
\end{figure}

To describe the continuum, we chose \texttt{blackbody + powerlaw} model modified \label{MandR}
by \texttt{Xspec} photoelectric absorption model \texttt{phabs}. After the
fitting procedure, we moved through the spectrum, adding cyclotron
absorption with cyclotron absorption model \texttt{cyclabs} with a step of the
central energy of $100$~eV and looked for the $\chi^2$ increase when setting the
depths of the harmonics to zero. This robust analysis revealed three spectra
from three different objects to have possible cyclotron absorptions. To
estimate the significances of absorption features, we followed the procedure
described in~\cite{Protassov02}, by running Monte Carlo simulations of
the spectra with the help of \texttt{Xspec} procedure \texttt{fakeit} and
calculating the percentage of the spectra which are fitted without
\texttt{cyclabs} component. The results are summarized in the Table~\ref{tab2}. 
The spectra of the magnetar candidate with the highest cyclotron feature significance 
detected are presented in figures \ref{fig1} and \ref{fig2}.
Using this procedure, we found the significances of all absorption features to
be below the margin $3\sigma$ detection limit.

\begin{table}
\label{tab2}
\begin{center}
\caption{The results of our analysis. In spite of the significant improvement of
the $\chi^2$ value, Monte Carlo simulations did not reveal significant
features.}
\begin{tabular}[!h]{lcccc}
\hline
\hline
Object & Obs. ID & Energy, keV & Width, eV & Significance, $\sigma$ \\
\hline
1E~1547.0-5408 & 0402910101 & 2.7 & 13 & 1.94 \\
1E~2259+586    & 0057540201 & 2.2 & 7  & 1.36 \\
XTE~J1810-197 & 0161360401 & 2.6 & 4  & 1.07 \\
\hline
\end{tabular}
\end{center}
\end{table}

\section*{Conclusions}
\indent \indent 
Though initial theoretical works~\cite{HoLai01,Zane01} have predicted a proton
cyclotron feature of equivalent width $0{.}7-0{.}75~\hbar\omega^{cycl}_{p}$,
in subsequent paper~\cite{LaiHo02} it has been shown that vacuum polarization
effects strongly suppress the cyclotron absorption feature, giving the
equivalent width about an order of magnitude lower. It should also be noted that
the theoretical calculations of the equivalent width of the cyclotron
feature were done for a local patch of the neutron star surface. Phase averaged
spectra, like those we are using here, would include contributions from various
magnetic field strengths, directions and effective temperatures, which would
further suppress the cyclotron feature. The results of our analysis indicate the
absence of the significant cyclotron absorptions in the quiescent spectra of the
magnetar candidates observed with \textsl{XMM-Newton}, what could be interpreted
due to combination of the two above-mentioned effects.

\section*{Acknowledgement}
\indent \indent M.~M. thanks to Lontkovskyi~D.~I. and Melnyk~G.~M. for helpful
and fruitful discussions. All the data analysis has been processed using the
computing facilities of Virtual Roentgen and Gamma-Ray Observatory in Ukraine
(\url{http://virgo.org.ua}). Both authors are grateful to McGill Pulsar Group
for the inappreciated information placed here:
\url{http://www.physics.mcgill.ca/~pulsar/magnetar/main.html}.

\newpage

\end{document}